\documentclass[aps,pra,twocolumn,showpacs,preprintnumbers,amsmath,amssymb]{revtex4-1}
\usepackage{soul}
\usepackage{amsmath}
\usepackage{esvect}

\usepackage{natbib}
\usepackage{graphicx}
\usepackage{color}

\usepackage{amssymb}

\usepackage[utf8]{inputenc}

\usepackage{float}

\usepackage{subfigure}

\usepackage{hyperref}

\begin{document}

\title{Analysis of the frequency shift in~coherent population trapping resonance's dynamic continuous-wave spectroscopy at the phase-jump modulation\\
and its comparison with the conventional approach}

\author{E.\,D.~Chivilis$^{1}$}
\affiliation{1. P.\,N. Lebedev Physical Institute of the Russian Academy of Sciences, Moscow, 119991 Russia}

\author{E.\,A.~Tsygankov$^{1}$}
\email[]{tsygankov.e.a@yandex.ru}
\affiliation{1. P.\,N. Lebedev Physical Institute of the Russian Academy of Sciences, Moscow, 119991 Russia}

\begin{abstract}
We present the research of~dynamic continuous-wave spectroscopy of~the coherent population trapping resonance at~the phase-jump modulation. $\Lambda$ system of~levels supplemented by~a~nonabsorbing state and bichromatic optical field, whose spectral components have different intensities, are~considered. We~demonstrate that the asymmetry leads to~an~additional nonlinear shift of~the error-signal frequency under unisotropic relaxation of~the ground-state density-matrix elements. We~also investigate the conventional approach where the frequency difference of~the optical field components is~harmonically modulated to~obtain the error signal. Comparison demonstrates that in~the high-frequency modulation regime the corresponding frequency shift is~more linear than at~the phase-jump modulation for nonshort integration times.
\end{abstract}

\maketitle

\section{Introduction}

The simplest way to~obtain the optical field to~induce the coherent population trapping resonance (CPT) is~the microwave current modulation of~the vertical-cavity surface-emitting laser (VCSEL)~\cite{affolderbach2000nonlinear}. This method does not require any bulky devices and therefore is~implemented in~chip-scale atomic clocks~\cite{8408999}. The modulation is~performed at~a~frequency $\Omega$ equal to~one-half of~the hyperfine frequency splitting of~an~alkali-metal atom ground state. The field's first sidebands are usually resonant since they are tuned to~an~alkali-metal atom D$_1$ line.

In the conventional approach, a~harmonic modulation with index $m$ and frequency $\omega_m$, \hbox{$\cos{\Omega t}\rightarrow\cos{(\Omega t+m\omega_m\cos{\omega_m t})}$} (in~the text, we~will refer to~this technique as~``modulation spectroscopy''), provides a~dispersively shaped response in~the optical field's absorption by~an~atomic ensemble. It~is~used as~the error signal to~lock the local oscillator's frequency to~that of~the ground-state splitting. In~the case of~the optically thin medium and equal powers of~the resonant sidebands, the error-signal zero and the CPT resonance minimum have the same frequency, which is~displaced by~the nonresonant light shift~\cite{1402-4896-93-11-114002}. But powers of~the VCSEL's first sidebands are generally unequal, which makes the CPT resonance asymmetric.
In~most cases, this produces an~additional shift of~the
error-signal frequency that depends on~modulation parameters~\cite{Phillips:05}. In~one of~our latest works, we~have demonstrated that this effect stems from the fact that the CPT resonance has a~multipeak structure~\cite{tsygankov2024nonlinear}. The error signal, in~its turn, is~not only a~single dispersive curve but a~group of~them. The side ones pull the frequency of~the central curve used for stabilization of~the local oscillator's frequency, wherein the shift nonlinearly depends on~the laser field intensity. It~is~known that techniques for suppression of~the CPT resonance frequency's light shift (in~what follows, we~use the term ``light shift'' for brevity) are based on~the assumption that the error-signal frequency linearly depends on~the laser field intensity~\cite{doi:10.1063/1.2360921,yudin2020general}. Therefore, the pulling reduces the effectiveness of~these approaches---the frequency can be~insensitive to~variations of~the optical field's power but be~displaced from the frequency of~the microwave transition unperturbed by~it.

The search for techniques with smaller sensitivity to~the resonance asymmetry and a~linear dependence of~the corresponding error-signal frequency on~the laser field intensity is~an~actual one. Dynamic continuous-wave spectroscopy of~the CPT resonance at~the phase-jump modulation (in~the text, we~will refer to~this technique as~``phase-jump spectroscopy''), proposed in~\cite{basalaev2019dynamic}, provides the error signal without the perturbation $m\cos{\omega_m t}$. The technique is~based on~the jump modulation of~the difference $2\varphi$ between phases of~the resonant sidebands: $2\varphi=\varphi_0$ for $t<0$, $2\varphi=\varphi_0+\Delta\varphi$ for $t\geq0$. Since the ground-state coherence induced by~the optical field depends on~$2\varphi$, the dark state prepared during $t<0$ becomes too some degree more ``bright'' for $t\geq0$. The transient process occurs to~rebuild the phase of~the ground-state coherence until it~does not become the dark state again.

The absorption during the transient process depends on~the ground-state coherence, which has even and odd terms over the two-photon detuning. The first one is~given by~the real part, which determines the shape of~the CPT resonance in~the steady-state; the second one is~given by~the imaginary part describing dispersion of~the two-photon transition. The sign of~the even part is~the same for forward and backward jumps and is~opposite for the odd term. Therefore, the difference in~absorption after forward and backward jumps leaves only the odd term, which can be~used as~the error signal. The question arises as~to~what extent the phase-jump spectroscopy~is~affected by~the CPT resonance asymmetry and whether the corresponding frequency shift is~a~nonlinear function of~the laser field intensity or~not, which we~analyze in~this work theoretically. 

\section{Phase-jump spectroscopy}

\begin{figure}[b]
\center{\includegraphics[width=\columnwidth]{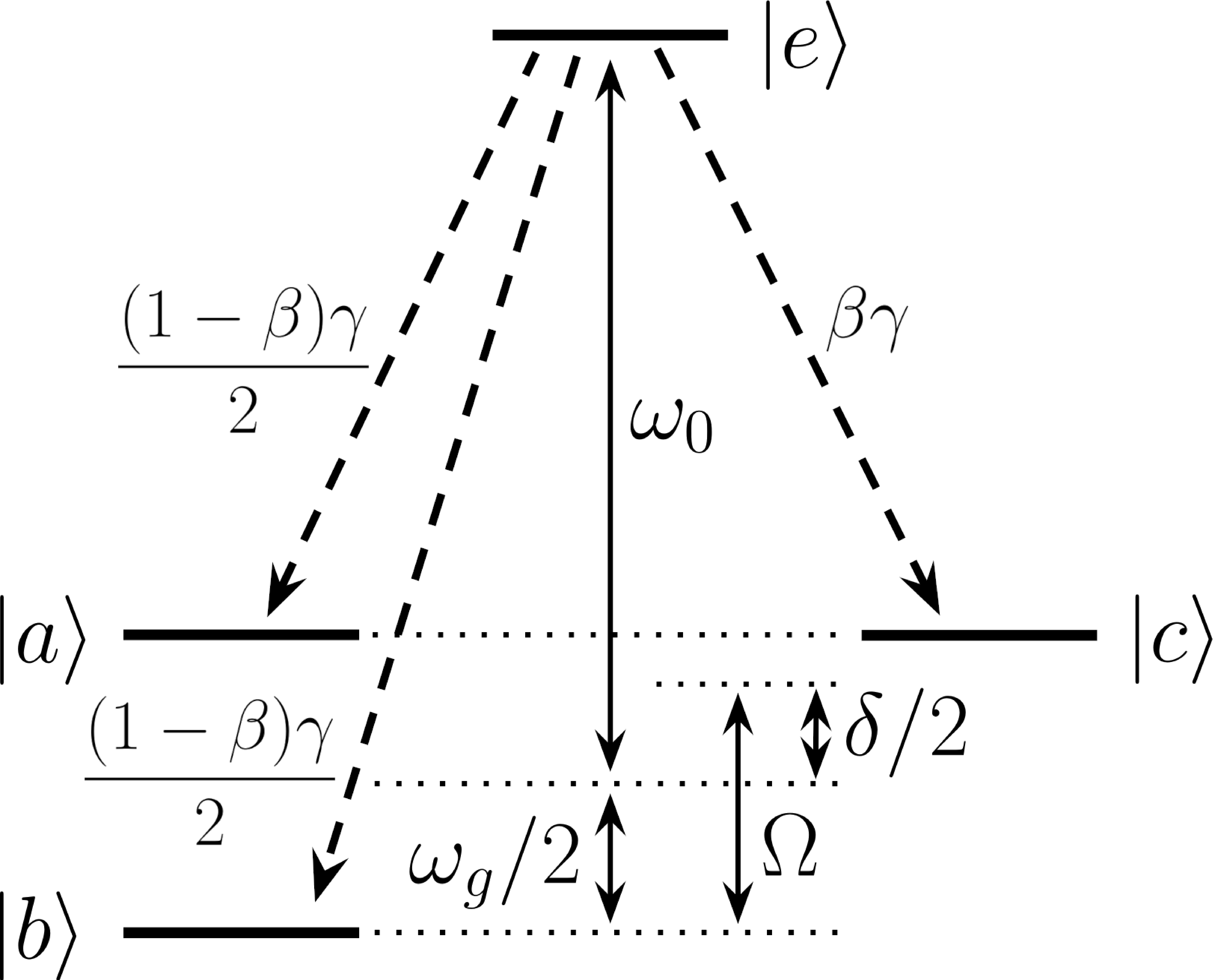}}
\caption{System of levels under consideration. Natural width of~the excited state is~$\gamma$. The spontaneous decay populates ground-state levels $|a\rangle$, $|b\rangle$, and $|c\rangle$ with rates $(1-\beta)\gamma/2$, $(1-\beta)\gamma/2$, and $\beta\gamma$, respectively.}
\label{LevelsScheme}
\end{figure}

We consider $\Lambda$ system of~levels with an~additional nonabsorbing state; see Fig.~\ref{LevelsScheme}. The optical field is~the following:
\begin{equation}
\begin{gathered}
\mathcal{E}(t)=e^{-i\omega_Lt}\left(\mathcal{E}_{-1}e^{i(\Omega t+\varphi)}+\mathcal{E}_1e^{-i(\Omega t+\varphi)}\right)/2+c.c.,
\end{gathered}
\label{OpticalField}
\end{equation}

\noindent where amplitudes are assumed to~be~real, $\mathcal{E}_{\mp1}=\mathcal{E}^*_{\mp1}$, wherein we~take dipole matrix elements as~$\langle e|\hat{d}|a\rangle=\langle a|\hat{d}|e\rangle=\langle e|\hat{d}|b\rangle=\langle b|\hat{d}|e\rangle=d$. The lower indices imply that the low-frequency and high-frequency components correspond to~sidebands $k=-1$ and $k=1$ of~the polychromatic VCSEL radiation used in~the experiment. The carrier frequency has detuning from half-sum of optical transitions; $\omega_L-\omega_0=\Delta_L$. The frequency difference between the optical field components has the small detuning $\delta$ from interval between ground-state levels: $2\Omega-\omega_g=\delta$, where $\delta\ll\omega_g$. The parameter $\varphi$ will be~used to~describe the phase-jump modulation. Here we~assume that $\Omega$ stands for the frequency of~the microwave modulation fed into the VCSEL to~obtain a~polychromatic radiation used to~induce the CPT resonance.

It is convenient to write the density matrix as
\begin{equation}
\hat{\rho}(t)=\begin{pmatrix}
\rho_{ee} & \rho_{ea} & \rho_{eb} & 0 \\
\rho_{ae} & \rho_{aa} & \rho_{ab} & 0 \\
\rho_{be} & \rho_{ba} & \rho_{bb} & 0 \\ 
0 & 0 & 0 & \rho_{cc} 
\end{pmatrix},
\end{equation}

\noindent where nondiagonal elements that are not induced by~the optical field were omitted right away. Then, we~have the Hamiltonian

\begin{equation}
\dfrac{\hat{\mathcal{H}}(t)}{\hbar}=\begin{pmatrix}
\omega_0 & V_{-1}(t) & V_1(t) & 0 \\
V^*_{-1}(t) & \omega_g/2 & 0 & 0 \\
V^*_1(t) & 0 & -\omega_g/2 & 0 \\ 
0 & 0 & 0 & \omega_g/2 
\end{pmatrix},
\end{equation}

\noindent where $V_{\mp1}(t)=V_{\mp1}e^{-i\left[(\omega_L\mp\Omega)t\mp\varphi\right]}$, $V_{\mp1}=d\mathcal{E}_{\mp1}/2\hbar$ are the Rabi frequencies. Here, we~have in~advance neglected by~the nonresonant terms in~the optical field. Also, we~assume that the low-frequency component induces transitions between states $|a\rangle$ and $|e\rangle$, while the high-frequency one couples states $|b\rangle$ and $|e\rangle$.
 
We will use the phenomenological constant $\Gamma$ to~describe relaxation of~the optical coherences. It~describes homogeneous broadening of~the optical line due to~collisions of~alkali-metal atoms with particles of~a~buffer gas. The relaxation of~the ground-state elements is~assumed to~be~unisotropic: the constant $\Gamma_c$ accounts for decay of~the ground-state coherence $\rho_{ab}$, while $\Gamma_g$ is~used to describe relaxation of the populations $\rho_{aa}$, $\rho_{bb}$, and $\rho_{cc}$ to~their equilibrium values $1/3$. Operating pressure of~buffer gases in~atomic cells is~several tens of~Torr, depending on~atomic cell size, therefore we~neglect the Doppler broadening as~long as~it~is~compared to~the homogeneous width of~the optical line. This also implies that $\gamma\ll\Gamma$. Also, the microwave Doppler effect is~canceled due to~the Dicke effect~\cite{PhysRev.89.472}. Then, the relaxation can be~written in~the form
{\footnotesize\begin{equation}
i\frac{\hat{\Gamma}\hat{\rho}}{\hbar}=\begin{pmatrix}
\gamma\rho_{ee} & \Gamma\rho_{ea} & \Gamma\rho_{eb} & 0 \\
\Gamma\rho_{ae} & \begin{gathered}\Gamma_g\left(\rho_{aa}-\frac13\right)\\
-\frac{(1-\beta)\gamma}{2}\rho_{ee}\end{gathered} & \Gamma_c\rho_{ab} & 0 \\
\Gamma\rho_{be} & \Gamma_c\rho_{ba} & \begin{gathered}\Gamma_g\left(\rho_{bb}-\frac13\right)\\
-\frac{(1-\beta)\gamma}{2}\rho_{ee}\end{gathered} & 0 \\ 
0 & 0 & 0 & \begin{gathered}\Gamma_g\left(\rho_{cc}-\frac13\right)\\
-\beta\gamma\rho_{ee}\end{gathered}
\end{pmatrix},
\end{equation}}

\noindent and incorporated in the quantum Liouville equation
\begin{equation}
i\hbar\dfrac{\partial}{\partial t}\hat{\rho}=[\hat{\mathcal{H}},\,\hat{\rho}]+\hat{\Gamma}\hat{\rho},
\end{equation}

\noindent where brackets denote the commutator.

We simplify initial equations for the density matrix elements using the following approximations. The low saturation regime is~considered, i.e., the excited-state population $\rho_{ee}$ is~neglected compared to~ground-state ones $\rho_{aa}$, $\rho_{bb}$, and $\rho_{cc}$. Further, we~introduce slowly-varying amplitudes and rapidly oscillating terms for coherences: $\rho_{ea}=\tilde{\rho}_{ea}e^{-i(\omega_L-\Omega)t+i\varphi}$, $\rho_{eb}=\tilde{\rho}_{eb}e^{-i(\omega_L+\Omega)t-i\varphi}$, and $\rho_{ab}=\tilde{\rho}_{ab}e^{-2i\Omega t}$ (such an~approach omits the nonresonant light shift, which is~out of~our interest as~far~as~it~is~a~linear function of~the laser field~intensity). This allows us~to adiabatically eliminate the excited-state and express optical coherences and $\rho_{ee}$ via the ground-state elements:
\begin{subequations}
\begin{equation}
\rho_{ee}=\dfrac{2\Gamma/\gamma}{\Delta^2_L+\Gamma^2}\left[V^2_{-1}\rho_{aa}+V^2_1\rho_{bb}+2V_{-1}V_1\text{Re}\left(e^{2i\varphi}\rho_{ab}\right)\right],
\end{equation}
\begin{equation}
\rho_{ea}=\dfrac{V_{-1}\rho_{aa}+e^{-2i\varphi}V_1\rho_{ba}}{\Delta_L+i\Gamma},
\end{equation}
\begin{equation}
\rho_{eb}=\dfrac{V_1\rho_{bb}+e^{2i\varphi}V_{-1}\rho_{ab}}{\Delta_L+i\Gamma}.
\end{equation}
\end{subequations}

Then, we use the obtained formulas to~derive system of~equations for the ground-state elements:
\begin{widetext}
\begin{subequations}
\begin{equation}
\begin{gathered}
\dfrac{\partial}{\partial t}\rho_{aa}=-(1+\beta)V^2_{-1}\dfrac{\Gamma}{\Delta^2_L+\Gamma^2}\rho_{aa}+(1-\beta)V^2_1\dfrac{\Gamma}{\Delta^2_L+\Gamma^2}\rho_{bb}\\
-2\beta V_{-1}V_1\dfrac{\Gamma}{\Delta^2_L+\Gamma^2}\text{Re}\left(e^{2i\varphi}\rho_{ab}\right)-2V_{-1}V_1\dfrac{\Delta_L}{\Delta^2_L+\Gamma^2}\text{Im}\left(e^{2i\varphi}\rho_{ab}\right)-\Gamma_g(\rho_{aa}-1/3),
\end{gathered}
\end{equation}
\begin{equation}
\begin{gathered}
\dfrac{\partial}{\partial t}\rho_{bb}=-(1+\beta)V^2_1\dfrac{\Gamma}{\Delta^2_L+\Gamma^2}\rho_{bb}+(1-\beta)V^2_{-1}\dfrac{\Gamma}{\Delta^2_L+\Gamma^2}\rho_{aa}\\
-2\beta V_{-1}V_1\dfrac{\Gamma}{\Delta^2_L+\Gamma^2}\text{Re}\left(e^{2i\varphi}\rho_{ab}\right)+2V_{-1}V_1\dfrac{\Delta_L}{\Delta^2_L+\Gamma^2}\text{Im}\left(e^{2i\varphi}\rho_{ab}\right)-\Gamma_g(\rho_{bb}-1/3),
\end{gathered}
\end{equation}
\begin{equation}
\begin{gathered}
\left\{i\dfrac{\partial}{\partial t}+\delta-(V^2_{-1}-V^2_1)\dfrac{\Delta_L}{\Delta^2_L+\Gamma^2}+i\left[\Gamma_c+\left(V^2_{-1}+V^2_1\right)\dfrac{\Gamma}{\Delta^2_L+\Gamma^2}\right]\right\}\rho_{ab}\\
=-e^{-2i\varphi}\dfrac{V_{-1}V_1}{\Delta^2_L+\Gamma^2}\left[i\Gamma(\rho_{aa}+\rho_{bb})-\Delta_L(\rho_{bb}-\rho_{aa})\right],
\end{gathered}
\label{rhoabEq}
\end{equation}
\label{initialequations}
\end{subequations}
\end{widetext}

\noindent where the equation for $\rho_{cc}$ is~not required due to~the conservation law $\text{Sp}(\hat{\rho})=1$. Eqs.~\eqref{initialequations} are used for the analysis of~the absorption after the phase jump. We~describe it~here as~follows. During the interval $t<0$ components of~the optical field~\eqref{OpticalField} have the phase difference $2\varphi$ and pump the system to~the steady-state. The corresponding quasi-stationary solution of~Eqs.~\eqref{initialequations} is~taken as~initial condition for $t\geq0$. At moment $t=0$, the phase jump occurs, that is, the phase difference $2\varphi$ instantly changes to zero: $2\varphi=0$ for $t\geq0$. The state formed before $t=0$ becomes greater absorbing since coherence $\rho_{ab}$ depends on~the value of $2\varphi$ and it~becomes to~a~some degree ``bright'' for the zero phase difference. Therefore, the transient process occurs, accompanied by~the absorption change until the system again reaches the steady-state regime. The described procedure is~valid for $t\gg1/\gamma$, i.e., for the interval when the excited-state elements have already reach their quasi-stationary state and have started to~adiabatically follow the ground state. Then, the absorption dynamics is~determined by~the transient solution for the ground-state density-matrix elements.

\subsection{Simplified case $\beta=0$}

Firstly, to~obtain some not too bulky analytical expressions, we~consider the case $\beta=0$, when no~optical pumping of~the nonabsorbing state $|c\rangle$ occurs. In~this case, from the structure of~Eqs.~\eqref{initialequations} follows that third-order differential equations (the number of~equations becomes smaller since the relation $\rho_{aa}+\rho_{bb}=2/3$ holds for $\beta=0$) for ground-state populations do~not depend on~the phase difference $2\varphi$. They are constant in~time and do~not affect the transient process when $2\varphi$ is~changed. This reflects the fact that, in~general, distribution of~populations does not depend on~the phase of~a~driving field. On~the contrary, the initial condition for $\rho_{ab}$ \hbox{depends~on $2\varphi$:}
\begin{subequations}
\begin{equation}
\rho_{ab}=-e^{-2i\varphi}\sqrt{\tilde{V}_{-1}\tilde{V}_1}\dfrac{2i/3-(\Delta_L/\Gamma)(\rho_{bb}-\rho_{aa})}{\tilde{\delta}+i\tilde{\Gamma}_c},
\label{initialcoherence}
\end{equation}
\begin{equation}
\rho_{bb}-\rho_{aa}=\dfrac23\dfrac{\tilde{V}_{-1}-\tilde{V}_1-4\tilde{\delta}(\Delta_L/\Gamma)\dfrac{\tilde{V}_{-1}\tilde{V}_1}{\tilde{\delta}^2+\tilde{\Gamma}^2_c}}{\tilde{\Gamma}_g+4\tilde{\Gamma}_c(\Delta_L/\Gamma)^2\dfrac{\tilde{V}_{-1}\tilde{V}_1}{\tilde{\delta}^2+\tilde{\Gamma}^2_c}},
\end{equation}
\label{initialcondition}
\end{subequations}

\noindent where we have introduced the following notations to~shorten the expressions: \hbox{$\tilde{V}_{\mp1}=V^2_{\mp1}\Gamma/(\Delta^2_L+\Gamma^2)$}, $\tilde{\Gamma}_c=\Gamma_c+\tilde{V}_{-1}+\tilde{V}_1$, $\tilde{\Gamma}_g=\Gamma_g+\tilde{V}_{-1}+\tilde{V}_1$, \hbox{$\tilde{\delta}=\delta-(V^2_{-1}-V^2_1)\Delta_L/(\Delta^2_L+\Gamma^2)$}, i.e., the detuning $\tilde{\delta}$ contains the resonant light shift, $\tilde{\Gamma}_c$ accounts for the power broadening, $\tilde{\Gamma}_g$ is~the increased relaxation rate of~populations due to~the optical pumping.

We note that the conservation law $\text{Sp}(\hat{\rho})=1$ always holds. The change of~absorption is~determined by~the change of~$\rho_{ee}$, which means that the ground-state populations should also change. However, in~the low saturation regime, $\rho_{ee}\ll\rho_{aa},\,\rho_{bb}$, and the change of~$\rho_{aa}$ and $\rho_{bb}$ can be~neglected. In~such a~case the transient process is~determined by the change of the ground-state coherence, which is~pointed out by~formulas~\eqref{initialcondition}. By~solving differential equation~(\ref{rhoabEq}) with initial condition~(\ref{initialcoherence}), we~obtain the following formula describing the difference in~the excited-state population $\Delta\rho_{ee}(t)$ for the forward jump ($2\varphi$ was~negative during $t<0$) and the backward jump ($2\varphi$ was~positive during $t<0$):
\begin{equation}
\begin{gathered}
\Delta\rho_{ee}(t)=\dfrac{8\sin{2\varphi}}{\gamma}\dfrac{\tilde{V}_{-1}\tilde{V}_1}{\tilde{\delta}^2+\tilde{\Gamma}^2_c}e^{-\tilde{\Gamma}_ct}\\
\cdot\left\{\left[2\tilde{\delta}/3+\tilde{\Gamma}_c(\Delta_L/\Gamma)(\rho_{bb}-\rho_{aa})\right]\cos{\tilde{\delta}t}\right.\\
\left.+\left[2\tilde{\Gamma}_c/3-\tilde{\delta}(\Delta_L/\Gamma)(\rho_{bb}-\rho_{aa})\right]\sin{\tilde{\delta}t}\right\},
\label{rhoeediff}
\end{gathered}
\end{equation}

As can be readily seen from formula~\eqref{rhoeediff}, $\Delta\rho_{ee}(t)$ is~not vanished in~the point $\tilde{\delta}=0$ because of~the term proportional to $(\Delta_L/\Gamma)(\rho_{bb}-\rho_{aa})$, i.e., when the CPT resonance is~asymmetric (neither even, nor odd function of~$\tilde{\delta}$).

The error signal is~defined as~the difference in~absorption for the forward and backward phase jump during the time interval $[0,\tau]$~\cite{basalaev2019dynamic}. The difference in~absorption during $[0,\tau]$ is~determined by the integral of Eq.~\eqref{rhoeediff} over $t$ multiplied by~$\gamma$. The subject of~our interest is~the shift of~the error signal zero due to~the CPT resonance asymmetry. We~obtain it~by~linearizing the solution over $\tilde{\delta}$ considering $\Delta_L/\Gamma\ll1$ (this inequality holds with a~wide margin for chip-scale atomic clocks) and finding a~value $\delta_{as}$ of~the two-photon detuning at~which it~is~equal to~the zero. It~reads~as
\begin{equation}
\delta_{as}(\tau)=-\dfrac{V^2_{-1}-V^2_1}{\Gamma}\dfrac{\Delta_L}{\Gamma}\dfrac{\tilde{\Gamma}_c}{\tilde{\Gamma}_g}\dfrac{1-e^{-\tilde{\Gamma}_c\tau}}{\left[2(1-e^{-\tilde{\Gamma}_c\tau})-e^{-\tilde{\Gamma}_c\tau}\tilde{\Gamma}_c\tau\right]}.
\label{shiftwithtau}
\end{equation}

This formula demonstrates that the shift associated with the asymmetry depends on~the integration time. In~particular, for $\tau\rightarrow0$ the last fraction in~Eq.~\eqref{shiftwithtau} is~equal to~one, and we~get
\begin{equation}
\delta_{as}(0)=-\dfrac{V^2_{-1}-V^2_1}{\Gamma}\dfrac{\Delta_L}{\Gamma}\dfrac{\Gamma_c+V^2_{-1}+V^2_1}{\Gamma_g+V^2_{-1}+V^2_1},
\end{equation}

\noindent where the last fraction written in~the explicit form shows that $\delta_{as}$ nonlinearly depends on~the laser field intensity since $\Gamma_c\neq\Gamma_g$. Further, we~should account for the resonant light shift to~obtain the value of~total displacement of~the error-signal zero, i.e., the equation $\tilde{\delta}=\delta-(V^2_{-1}-V^2_1)\Delta_L/(\Delta^2_L+\Gamma^2)=\delta_{as}(0)$ should be~solved. Then,
\begin{equation}
\begin{gathered}
\delta_{tot}(0)=\dfrac{V^2_{-1}-V^2_1}{\Gamma}\dfrac{\Delta_L}{\Gamma}-\dfrac{V^2_{-1}-V^2_1}{\Gamma}\dfrac{\Delta_L}{\Gamma}\dfrac{\tilde{\Gamma}_c}{\tilde{\Gamma}_g}\\
=-\dfrac{V^2_{-1}-V^2_1}{\Gamma}\dfrac{\Delta_L}{\Gamma}\dfrac{\Gamma_c-\Gamma_g}{\Gamma_g+(V^2_{-1}+V^2_1)/\Gamma}.
\label{shiftzerotau}
\end{gathered}
\end{equation}

As~can be~seen, there is~self-compensation of~the resonant light shift by~the one due to~the asymmetry when $\Gamma_c=\Gamma_g$, i.e., when there is~the isotropic relaxation. In~the vast majority of~experimental cases $\Gamma_c\neq\Gamma_g$, $\delta_{tot}\neq0$ and it~also nonlinearly depends on~the optical field intensity.

The final result for $\tau\rightarrow\infty$ (all transient process is~integrated) read~as
\begin{equation}
\delta_{tot}(\infty)=-\dfrac12\dfrac{V^2_{-1}-V^2_1}{\Gamma}\dfrac{\Delta_L}{\Gamma}\dfrac{\Gamma_c-2\Gamma_g-(V^2_{-1}+V^2_1)/\Gamma}{\Gamma_g+(V^2_{-1}+V^2_1)/\Gamma}.
\label{shiftinfinitetau}
\end{equation}

\begin{figure}[b]
\center{\includegraphics[width=\columnwidth]{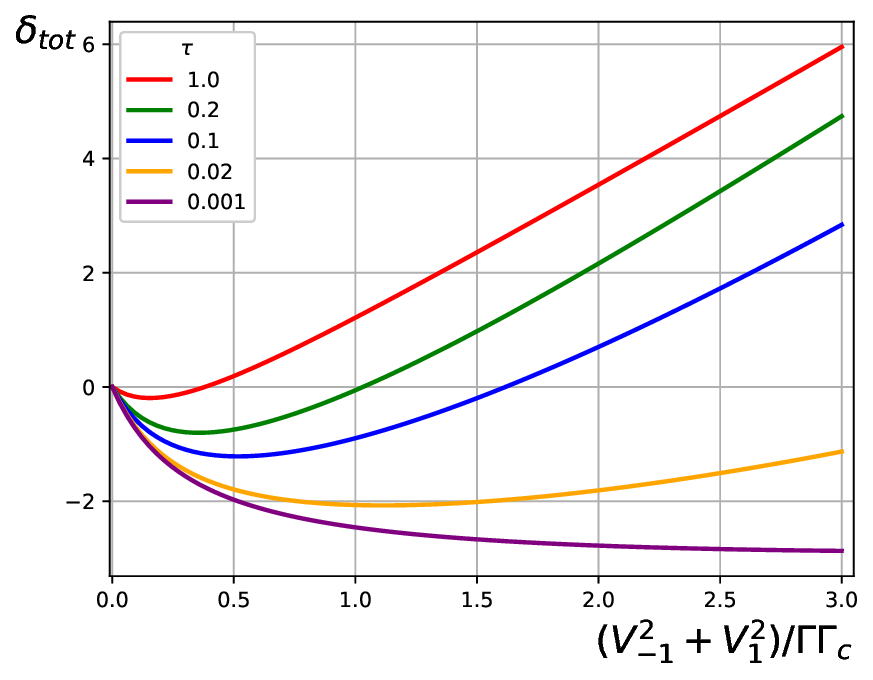}}
\caption{Frequency shift of~the error-signal zero for different integration times plotted via~\eqref{shiftwithtau}. Units on~the horizontal axis are the ratio of~the power broadening to~the relaxation rate of~the ground-state coherence. The vertical axis is~given in~units $\left[(V^2_{-1}-V^2_1)/\Gamma\right]\Delta_L/\Gamma$. The ratio of~the coherence decay rate to~that of~ground-state populations to~their equilibrium values $1/3$ is~$\Gamma_c/\Gamma_g=3$.}
\label{delta_as}
\end{figure}

Fig.~\ref{delta_as} demonstrates $\delta_{tot}(\tau)$ for different integration times. As~can be seen, the shift becomes more nonlinear for greater $\tau$. When the power broadening dominates over the relaxation rate, $(V^2_{-1}+V^2_1)/\Gamma\gg\Gamma_{g,c}$, $\delta_{tot}(\infty)$ is~proportional to~the optical field intensity, while $\delta_{tot}(0)$ does not depend on~it~at all. It~seems that it~will be~enough to~use radiation of~a~sufficiently high power to~solve the problem. But this condition is~not easy feasible in~the case of~chip-scale atomic clocks---the maximal VCSEL optical power commonly is~limited by~several hundreds of~$\mu$W, providing nearly ten-fold excess of~the power broadening over the ground-state coherence relaxation rate. However, an~increase in~the power broadening leads to~a~less steeper error signal and decrease in~the short-term frequency stability.

\subsection{Accounting for optical pumping of the nonabsorbing state $|c\rangle$}

\begin{figure}[b]
\center{\includegraphics[width=\columnwidth]{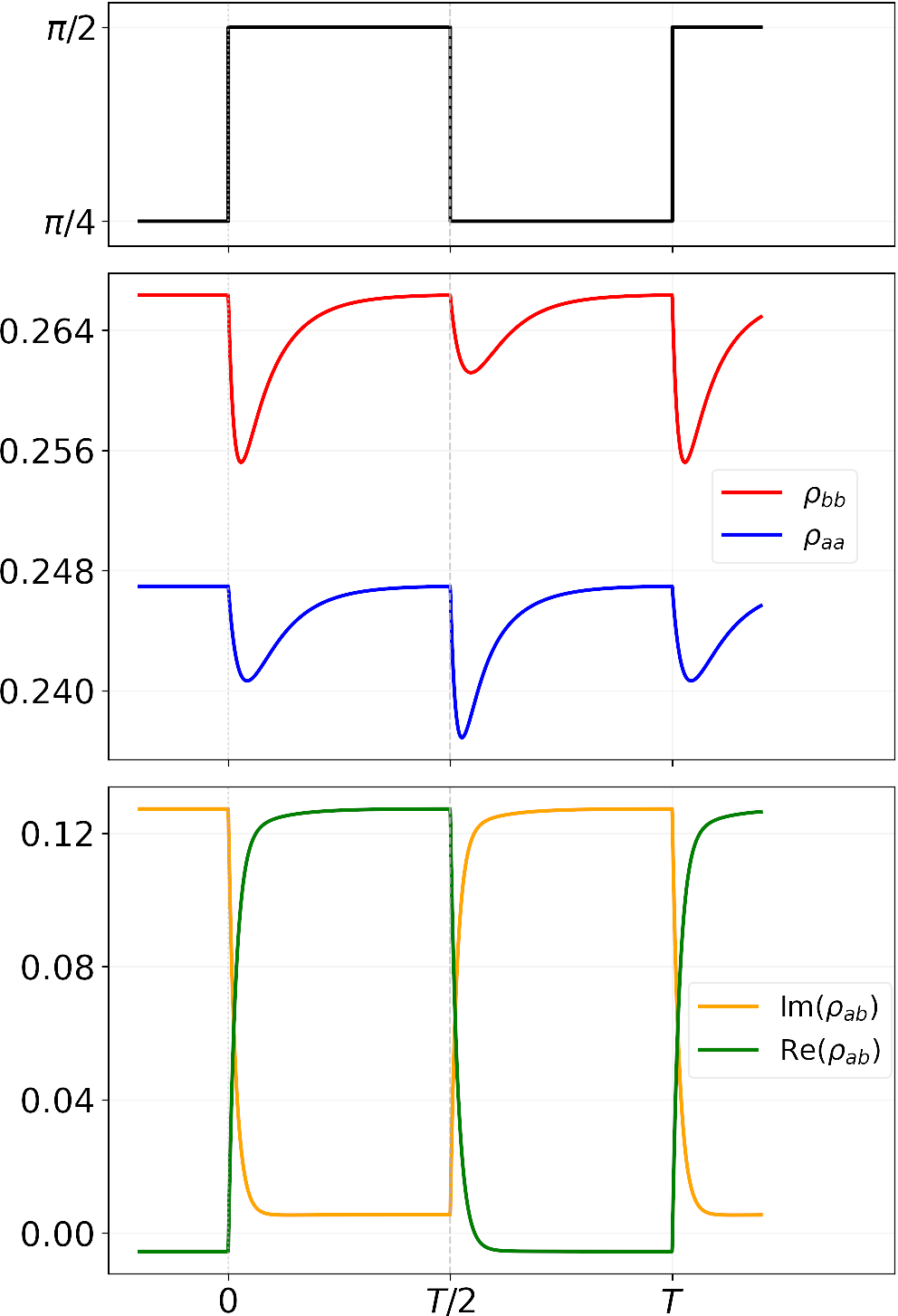}}
\caption{Dynamics of~the ground-state density-matrix elements after forward and backward phase jumps. The values of~parameters are $\beta=0.2$, $\delta/\tilde\Gamma_c=0.21$, $\Gamma_c/\Gamma_g=3.0$, $\Delta_L/\Gamma=0.1$, $(V^2_{-1}+V^2_{1})/(\Gamma\Gamma_c)=1.0$, ${V^2_{1}/V^2_{-1}}=0.9$. The period of~modulation is~long enough to~ensure the steady-state regime before the following jump.}
\label{Dynamics}
\end{figure}

In the case $\beta\neq0$ the relation $\rho_{aa}+\rho_{bb}=const$ no~longer holds. After a~phase jump, the superposition of~states $|a\rangle$ and $|b\rangle$ become more ``bright'' leading to~a~decrease in~populations of~absorbing sublevels and optical pumping of~the state $|c\rangle$. After a~some time, when coherence is~sufficiently rebuild~by~the optical field, atoms begin to~accumulate at~states $|a\rangle$ and $|b\rangle$: their relaxation from the state $|c\rangle$ to~absorbing sublevels leads to~trapping of~population in~coherent superposition of states $|a\rangle$ and $|b\rangle$. We~demonstrate it~by~presenting the curves given in~Fig.~\ref{Dynamics}.

We~investigated the case $\tau\rightarrow0$. The transient processes were integrated in~the interval $0,\,1/1000\,\tilde{\Gamma}_c$. After that, we~found the value of~$\delta$ providing zero of~the error signal. Fig.~$3$ demonstrates the obtained result. For $\beta=0$ the numerical result coincides with the one given by~formula~\eqref{shiftzerotau}. For longer integration times, dependence of~the shift on~the power broadening is~between cases $\tau\rightarrow0$ and $\tau\rightarrow\infty$.

When the optical pumping of~the state $|c\rangle$ occurs, the shift becomes sensitive to~the laser field intensity in~the region $(\tilde{V}_{-1}+\tilde{V}_1)/\Gamma_c$. The proportionality coefficient grows with $\beta$. However, the change in~shift is~smaller than $10^{-4}\tilde{\Gamma}_c$. Therefore, there is~only the decrease in~the slope for short integration times like $1/1000\tilde{\Gamma}_c$: the steepness of~the error signal is~$\sim50$ times smaller compared to~$\tau\rightarrow\infty$.

\section{Conventional modulation spectroscopy}

Here we~investigate the case with harmonic sinusoidal modulation of~the difference between~spectral components phases, namely, $2\varphi=2m\sin{\omega_mt}$, where $m$ is~the phase modulation index and $\omega_m$ is~the modulation frequency. This modulation provides oscillations of~the light absorption at~multiples of~$\omega_m$~\cite{1402-4896-93-11-114002}. Odd harmonics have a~dispersive shape and can be~used in~feedback for stabilization of~the local oscillator's frequency. The first one is~usually used as~far as~it~has the greatest slope. The corresponding oscillations can be~represented as~the sum of~the in-phase and quadrature components, \hbox{$\kappa_1(t)=A_I\cos{\omega_mt}+A_Q\sin{\omega_mt}$}. In~\cite{tsygankov2024nonlinear}, we~have demonstrated that the modulation provides a~group of~dispersively shaped curves for the quadrature signal. In~the case of~the asymmetry, side curves pull the frequency of~the central one, which can be~used as~the error signal. Curves become resolved at~a~high enough frequency $\omega_m$, leading to~suppression of~the pulling. In~this subsection we~will derive an~analytical expression for the shift of~the central curve frequency in~order to~understand, wether it~is~nonlinear over the laser field intensity or~not in~the case $\Gamma_c\neq\Gamma_g$ and presence of~the state $|c\rangle$ (under the isotropic relaxation and for $\beta=0$, the shift is~a~linear function~\cite{tsygankov2024nonlinear}).

\subsection{Absence of the state $|c\rangle$}

At first, it is convenient to make the substitution \hbox{$\tilde{\rho}_{ab}=\bar{\rho}_{ab}e^{-2im\sin{\omega_mt}}$}. Secondly, as~far as~for $\beta=0$ the relation $\rho_{aa}+\rho_{bb}=2/3$ holds, we~replace $\rho_{bb}$ by~$2/3-\rho_{aa}$. Also, in~the case of~the conventional modulation, populations and coherences oscillate at~multiples of~$\omega_m$ frequency, which we~will account for~by the Fourier series expansion. In~this situation, it~is~convenient to~introduce the difference from the equilibrium value $1/3$ for the population of~the state $|a\rangle$, $\rho_{aa}=\alpha+1/3$. This gives
\begin{subequations}
\begin{equation}
\rho_{ee}=\dfrac{2}{\gamma}\left[\dfrac{\tilde{V}_{-1}+\tilde{V}_1}{3}+\alpha(\tilde{V}_{-1}-\tilde{V}_1)+2\sqrt{\tilde{V}_{-1}\tilde{V}_1}\text{Re}(\bar{\rho}_{ab})\right],
\end{equation}
\begin{equation}
\dfrac{\partial}{\partial t}\alpha=-\dfrac{\tilde{V}_{-1}-\tilde{V}_1}{3}-\tilde{\Gamma}_g\alpha-2\sqrt{\tilde{V}_{-1}\tilde{V}_1}\dfrac{\Delta_L}{\Gamma}\text{Im}(\bar{\rho}_{ab}),
\end{equation}
\begin{equation}
\begin{gathered}
\left(i\dfrac{\partial}{\partial t}+\tilde{\delta}+2m\omega_m\cos{\omega_mt}+i\tilde{\Gamma}_c\right)\bar{\rho}_{ab}=\\
-2\sqrt{\tilde{V}_{-1}\tilde{V}_1}\left(\dfrac i3+\dfrac{\Delta_L}{\Gamma}\alpha\right).
\end{gathered}
\end{equation}
\end{subequations}

Further, we make Fourier series expansion of~the density matrix elements over the frequency $\omega_m$: \hbox{$\rho_{ee}=\sum^{\infty}_{k=-\infty}E_ke^{-ik\omega_mt}$}, $\alpha=\sum^{\infty}_{k=-\infty}G_ke^{-ik\omega_mt}$, \hbox{$\bar{\rho}_{ab}=\sum^{\infty}_{k=-\infty}C_ke^{-ik\omega_mt}$}. We~are interested in~the central dispersive curve of~the quadrature signal. Its slope is~maximized at~$m\simeq0.54$ due to~the proportionality factor $J_0(2m)J_1(2m)$~\cite{tsygankov2024nonlinear}. Therefore, the approximation of~small modulation index $m\ll1$ is~quite good for the analysis and derived expressions will work for $2m\lesssim1$. Then, only Fourier amplitudes with indices $\mp1$ should be~hold in~the expansion.

From the fact that populations are real functions follows that $E_{-1}=E^*_1$, $G_{-1}=G^*_1$, which allows us~to~write amplitudes of~the in-phase and quadrature signals~as
\begin{subequations}
\begin{equation}
A_I=\dfrac{4}{\gamma}\left[(\tilde{V}_{-1}-\tilde{V}_1)\text{Re}(G_1)+\sqrt{\tilde{V}_{-1}\tilde{V}_1}\text{Re}(C_{-1}+C_1)\right],
\end{equation}
\begin{equation}
A_Q=\dfrac{4}{\gamma}\left[(\tilde{V}_{-1}-\tilde{V}_1)\text{Im}(G_1)-\sqrt{\tilde{V}_{-1}\tilde{V}_1}\text{Im}(C_{-1}-C_1)\right].
\end{equation}
\end{subequations}

In its turn, the equations for ground-state Fourier amplitudes have the following form:
\begin{subequations}
\begin{equation}
\tilde{\Gamma}_gG_0=-\dfrac{\tilde{V}_{-1}-\tilde{V}_1}{3}+i\sqrt{\tilde{V}_{-1}\tilde{V}_1}\dfrac{\Delta_L}{\Gamma}(C_0-C^*_0),
\end{equation}
\begin{equation}
(\tilde{\Gamma}_g-i\omega_m)G_1=i\sqrt{\tilde{V}_{-1}\tilde{V}_1}\dfrac{\Delta_L}{\Gamma}(C_1-C^*_{-1}),
\end{equation}
\begin{equation}
\begin{gathered}
(\tilde{\delta}+i\tilde{\Gamma}_c)C_0+m\omega_m(C_1+C_{-1})=\\
-2\sqrt{\tilde{V}_{-1}\tilde{V}_1}\left(\dfrac i3+\dfrac{\Delta_L}{\Gamma}G_0\right),
\end{gathered}
\end{equation}
\begin{equation}
(\tilde{\delta}+\omega_m+i\tilde{\Gamma}_c)C_1 +m\omega_mC_0=-2\sqrt{\tilde{V}_{-1}\tilde{V}_1}\dfrac{\Delta_L}{\Gamma}G_1,
\end{equation}
\begin{equation}
(\tilde{\delta}-\omega_m+i\tilde{\Gamma}_c)C_{-1} + m\omega_mC_0=-2\sqrt{\tilde{V}_{-1}\tilde{V}_1}\dfrac{\Delta_L}{\Gamma}G^*_1.
\end{equation}
\end{subequations}

Finally, we~use the obtained equations to~derive the frequency shift of~the resolved ($\omega_m\gg\tilde{\Gamma}_c$) central dispersive curve, which reads~as
\begin{equation}
\delta^{Q}_{as}=-\dfrac{V^2_{-1}-V^2_1}{\Gamma}\dfrac{\Delta_L}{\Gamma}\dfrac{\tilde{\Gamma}_c}{\tilde{\Gamma}_g}
\end{equation}

\noindent under linearization over $m$ and $\Delta_L/\Gamma$. The formula demonstrates that the shift is~the same as~for the phase-jump modulation in~the case $\tau\rightarrow0$.
This result can be~treated in~the way that both techniques probe the resonance with the same asymmetry when the multipeak structure is~resolved.
\begin{figure}[t]
\center{\includegraphics[width=\columnwidth]{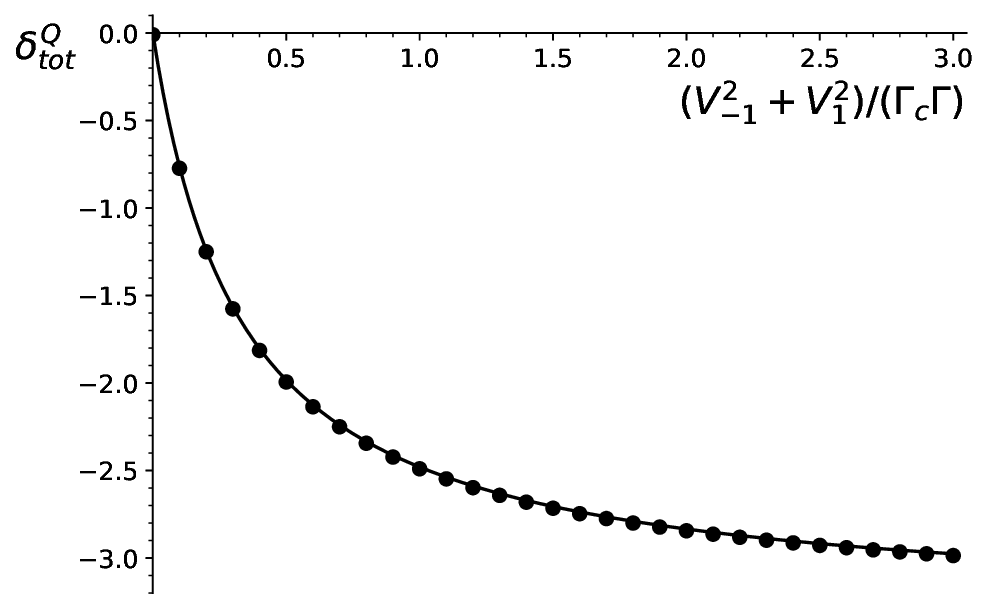}}
\caption{The frequency shift of~the central dispersive curve's zero. The solid line is~plotted via formula~\eqref{shiftzerotau}, the circles are~numerical calculations for $\beta=0.5$. The values of~other parameters are the same as~in Fig.~\ref{Dynamics}. The shape of~the curve coincides with that for the phase-jump technique at~integration time of~$1/1000\tilde{\Gamma}_c$.}
\label{QuadrShift}
\end{figure}
\begin{figure}[b]
\center{\includegraphics[width=\columnwidth]{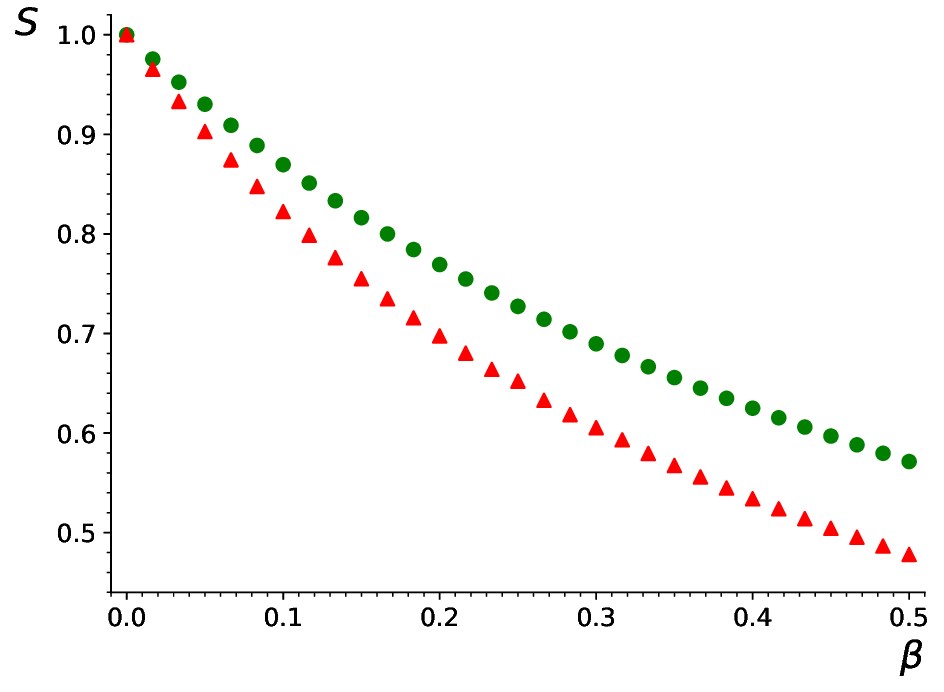}}
\caption{Dependence of~the normalized slope on~$\beta$ for the phase-jump technique (circles) and the conventional approach (triangles). The values of~parameters are \hbox{$\tilde{V}_{-1}+\tilde{V}_1=\Gamma_c$}, $\tilde{V}_{-1}=\tilde{V}_1$, $\Delta_L=0$, $\Gamma_g=\Gamma_c/3$. The integration time for phase jumps is~$1/1000\tilde{\Gamma}_c$. For harmonic modulation, $\omega_m=100\tilde{\Gamma}_c$, $m=0.54$.}
\label{Steepness}
\end{figure}

\subsection{General case of~$\beta\neq0$}

The system was supplemented by~the equation for $\rho_{bb}$ as~far as~$\rho_{cc}\neq1/3$~for $\beta\neq0$. For numerical investigation we~set \hbox{$\omega_m=100\,\tilde{\Gamma}_c$} to~well-resolve the multi-dispersive structure of~the quadrature signal~\cite{tsygankov2024nonlinear}. The solution was obtained for \hbox{$t>10/(\Gamma_c+\tilde{V}_{-1}+\tilde{V}_1)$} to~ensure the steady-state regime. We~have obtained dependence of~the frequency shift on~the power broadening for different values of~$\beta$ from $0.1$ to $0.5$. The calculations demonstrated that there is~no~difference in~the shift value within the limits of~computational accuracy; see Fig.~\ref{QuadrShift}.

The only difference between techniques is~in~the decrease of~the error-signal steepness with $\beta$. Fig.~\ref{Steepness} demonstrates comparison for a~symmetric resonance. The difference in~slope is~slightly in~favor of~the phase-jump technique. For $\beta=0.1,\,0.3,\,0.5$ the steepness is~greater by~$\sim5,\,9,\,11\%$, respectively. However, the correct comparison should account for the noise level, which should be~done experimentally as~far as~the techniques use different principles of~the detection.

\section{Conclusion}

Our analysis has demonstrated that the phase-jump modulation does not provide a~smaller frequency shift of~the error-signal zero compared to~the conventional approach with harmonic modulation. For~both techniques, it~arises at~unequal relaxation rates of~the ground-state populations and coherence. For a~specific case of~their equality, the self-compensation effect takes place: the resonant light shift is~opposite to~the displacement due to~the CPT resonance asymmetry. Considering the phase-jump spectroscopy, the shift is~more nonlinear over the optical field intensity for nonshort integration times. Accounting for the nonabsorbing sublevel has demonstrated that the steepness of~the error signal drops slightly slower with the branching coefficient for the phase-jump technique. On~the other hand, since the detection is~noncontinuous, it~is~affected by~the Dick effect. Additionally, while the phase jumps operate in~the window of~low-frequency noise, the harmonic modulation can be~implemented at~relatively high frequencies to~suppress $1/f$ noise. Therefore, our conclusion is~that the conventional approach is~more suitable for the light shift suppression techniques than the phase-jump spectroscopy.

\section{Acknowledgments}

This work was financed by~funds from the state assignment to~the P.\,N. Lebedev Physical Institute of~the Russian Academy of~Sciences. No~additional grants were received for the implementation or~guidance of~this research.

\bibliographystyle{apsrev4-1}
\bibliography{references}

@article{doi:10.1063/1.2360921,
author = {Shah, V. and Gerginov, V. and Schwindt, P. D. D. and Knappe, S. and Hollberg, L. and Kitching, J.},
title = {Continuous light-shift correction in modulated coherent population trapping clocks},
journal = {Applied Physics Letters},
volume = {89},
number = {15},
pages = {151124},
year = {2006},
doi = {10.1063/1.2360921},
URL = {https://doi.org/10.1063/1.2360921},
eprint = {https://doi.org/10.1063/1.2360921}
}

@article{1402-4896-93-11-114002,
  author={Chuchelov, D. S. and Vassiliev, V. V. and Vaskovskaya, M. I. and Velichansky, V. L. and Tsygankov, E. A. and Zibrov, S. A. and Petropavlovsky, S. V. and Yakovlev, V. P.},
  title={Modulation spectroscopy of coherent population trapping resonance and light shifts},
  journal={Physica Scripta},
  volume={93},
  number={11},
  pages={114002},
  url={http://stacks.iop.org/1402-4896/93/i=11/a=114002},
  year={2018}
}

@article{Phillips:05,
author = {David F. Phillips and Irina Novikova and Christine Y.-T. Wang and Ronald L. Walsworth and Michael Crescimanno},
journal = {J. Opt. Soc. Am. B},
number = {2},
pages = {305--310},
publisher = {OSA},
title = {Modulation-induced frequency shifts in a coherent-population-trapping-based atomic            clock},
volume = {22},
month = {Feb},
year = {2005},
url = {http://josab.osa.org/abstract.cfm?URI=josab-22-2-305},
doi = {10.1364/JOSAB.22.000305},
}

@article{PhysRev.89.472,
  title = {The Effect of Collisions upon the Doppler Width of Spectral Lines},
  author = {Dicke, R. H.},
  journal = {Phys. Rev.},
  volume = {89},
  issue = {2},
  pages = {472--473},
  numpages = {0},
  year = {1953},
  month = {Jan},
  publisher = {American Physical Society},
  doi = {10.1103/PhysRev.89.472},
  url = {https://link.aps.org/doi/10.1103/PhysRev.89.472}
}

@misc{basalaev2019dynamic,
    title={Dynamic continuous-wave spectroscopy of coherent population trapping at phase-jump modulation},
    author={M. Yu. Basalaev and V. I. Yudin and A. V. Taichenachev and M. I. Vaskovskaya and D. S. Chuchelov and S. A. Zibrov and V. V. Vassiliev and V. L. Velichansky},
    year={2019},
    eprint={1911.00575},
    archivePrefix={arXiv},
    primaryClass={physics.atom-ph}
}

@article{tsygankov2024nonlinear,
  title={Nonlinear frequency shift caused by asymmetry of the multipeak coherent population trapping resonance},
  author={Tsygankov, EA and Chuchelov, DS and Vaskovskaya, MI and Vassiliev, VV and Zibrov, SA and Velichansky, VL},
  journal={Physical Review A},
  volume={109},
  number={5},
  pages={053703},
  year={2024},
  publisher={APS}
}

@INPROCEEDINGS{8408999,
  author={Cash, Peter and Krzewick, Will and Machado, Paul and Overstreet, K. Richard and Silveira, Mike and Stanczyk, Matt and Taylor, Dwayne and Zhang, Xianli},
  booktitle={2018 European Frequency and Time Forum (EFTF)}, 
  title={Microsemi Chip Scale Atomic Clock (CSAC) technical status, applications, and future plans}, 
  year={2018},
  volume={},
  number={},
  pages={65-71},
  doi={10.1109/EFTF.2018.8408999}}

@article{affolderbach2000nonlinear,
  title={Nonlinear spectroscopy with a vertical-cavity surface-emitting laser (VCSEL)},
  author={Affolderbach, C and Nagel, A and Knappe, S and Jung, C and Wiedenmann, D and Wynands, R},
  journal={Applied Physics B},
  volume={70},
  number={3},
  pages={407--413},
  year={2000},
  publisher={Springer}
}

@article{yudin2020general,
  title={General methods for suppressing the light shift in atomic clocks using power modulation},
  author={Yudin, VI and Basalaev, M Yu and Taichenachev, AV and Pollock, JW and Newman, ZL and Shuker, Moshe and Hansen, Azure and Hummon, MT and Boudot, Rodolphe and Donley, Elizabeth A and others},
  journal={Physical Review Applied},
  volume={14},
  number={2},
  pages={024001},
  year={2020},
  publisher={APS}
}

\end{document}